\documentstyle[12pt,aasms4]{article}

\def\d{SN~1997D}
\def\kms{km s$^{-1}$}
\def\Ha{H$\alpha$}
\def\Hb{H$\beta$}

\def\Mv{$M_{\rm V}$}

\def\bv{(B$-$V)}
\def\M{$M_{\odot}$}
\def\R{$R_{\odot}$}
\def\m100{mag/100$^d$}
\def\ni{{$^{56}$Ni}}
\def\c57{{$^{57}$Co}\/}
\def\co{{$^{56}$Co}}

\def\ti44{{$^{44}$Ti}\/}
\def\r0{{$R_0$}}
\def\ltsima{$\; \buildrel < \over \sim \;$}
\def\ltsim{\lower.5ex\hbox{\ltsima}}
\def\gtsima{$\; \buildrel > \over \sim \;$}
\def\gtsim{\lower.5ex\hbox{\gtsima}}

\begin{document}
\lefthead{Turatto et al.}
\righthead{Supernova 1997~D: a case for very low $^{56}$Ni mass}

\title{The peculiar Type II Supernova 1997~D: 
A case for a very low $^{56}$Ni mass
\footnote[0]{Based on observations collected at ESO-La Silla and
Cerro Tololo 
InterAmerican Observatory (Chile). Cerro Tololo Inter-American
Observatory, National Optical Astronomy Observatories, operated by
the
Association of Universities for Research in Astronomy, Inc., (AURA),
under cooperative agreement with the National Science Foundation}}

\author{M. Turatto\altaffilmark{1,2,3}, P.A.
Mazzali\altaffilmark{4,5,3},
T.R. Young\altaffilmark{6,4}, K. Nomoto\altaffilmark{6,3,7}, 
K. Iwamoto\altaffilmark{6,3}}

\author{S. Benetti\altaffilmark{1}, E. Cappellaro\altaffilmark{2,3},  
I.J. Danziger\altaffilmark{5}, D.F. de Mello\altaffilmark{8}, 
M.M. Phillips\altaffilmark{9}, N.B. Suntzeff\altaffilmark{9},
A. Clocchiatti\altaffilmark{9,10}, A. Piemonte\altaffilmark{1}, 
B. Leibundgut\altaffilmark{11}, R. Covarrubias\altaffilmark{9}, 
J. Maza\altaffilmark{12}, J. Sollerman\altaffilmark{13},
}

\altaffiltext{1}{European Southern Observatory, Alonso de Cordova
3107,
Vitacura, Casilla 19001 Santiago 19, Chile}
\altaffiltext{2}{Osservatorio Astronomico di Padova, vicolo 
dell'Osservatorio 5, I-35122 Padova, Italy}
\altaffiltext{3}{Institute for Theoretical Physics, University of
California,
Santa Barbara, CA, USA}
\altaffiltext{4}{National Astronomical Observatory, Mitaka, Tokyo
181, Japan}
\altaffiltext{5}{Osservatorio Astronomico di Trieste, via G.B.
Tiepolo 11, 
I-34131 Trieste, Italy}
\altaffiltext{6}{Department of Astronomy, University of Tokyo,
Bunkyo--ku, 
Tokyo 113, Japan}
\altaffiltext{7}{Research Center for the Early Universe,  
University of Tokyo, Bunkyo--ku, Tokyo 113, Japan}
\altaffiltext{8}{Space telescope Science Institute, 3700 San Martin
Drive,
Baltimore, MD 21218, USA}
\altaffiltext{9}{Cerro Tololo Interamerican Observatory, Casilla 603, 
La Serena, Chile}
\altaffiltext{10}{Departamento de Astronomia, 
P. Universitad Catolica, Casilla 104, Santiago 22, Chile}
\altaffiltext{11}{European Southern Observatory,
Karl-Schwarzschild-Strasse 2,
D-85748 Garching bei M\"unchen, Germany}
\altaffiltext{12}{Departamento de Astronomia, Universidad de Chile, 
Casilla 36-D, Santiago, Chile}
\altaffiltext{13}{Stockholm Observatory, SE--133 36 Saltj\"obaden,
Sweden}

\begin{abstract}

\d\/ in NGC 1536 is possibly the least luminous and energetic Type II 
supernova discovered to date.  
The entire light curve is subluminous, never reaching $M_V = -14.65$. 
The radioactive tail follows the \co\ decay slope. 
In the case of nearly complete trapping of the $\gamma$-rays, the
\ni\ mass 
derived from the tail brightness is extremely small, $\sim 0.002$ \M.  
At discovery the spectra showed a red continuum and line velocities
of 
the order of 1000 \kms. 
The luminosity and the photospheric expansion velocity suggest that
the 
explosion occurred about 50 days before discovery, and that a plateau 
probably followed. 
Model light curves and spectra of the explosion of a 26 \M\ star 
successfully fit the observations.  
Low mass models are inconsistent with the observations. 
The radius of the progenitor, constrained by the prediscovery upper
limits, is \r0\ \ltsim 300 \R.  
A low explosion energy of $\sim 4 \times 10^{50}$ ergs is then
required 
in the modeling. 
The strong \ion{Ba}{2} lines in the photospheric spectra are
reproduced 
with a solar abundance and low $T_{\rm eff}$.
A scenario in which the low \ni\/ mass observed in \d\/ is due to 
fall--back of material onto the collapsed remnant of the explosion of
a 
25--40 \M\/ star appears to be favored over the case of the explosion
of 
an 8--10 \M\ star with low \ni\/ production.
\end{abstract}

\keywords{nucleosynthesis -- stars: evolution -- 
supernovae: general ---  supernovae: individual (SN 1997D)
}
 
\section{Introduction}

\d\/ was serendipitously discovered on Jan. 14.15 U.T.
(\cite{demello}) 
in NGC~1536, a morphologically disturbed spiral galaxy belonging to a
high density group (\cite{maia}).  Although NGC~1536 
($v_{\rm helio}=1296$ \kms, RC3) is one of the galaxies patrolled
visually 
by Rev. Evans for his SN search (private communication), he missed
the 
SN because its brightness apparently never exceeded his detection
limit.

The first spectrum showed the main features of type II SNe, but also 
revealed peculiarities which make this object unique.  
Particularly noteworthy were the extremely slow expansion velocities, 
the red color and the strong \ion{Ba}{2} lines (\cite{tur98}). 
A campaign of photometric and spectroscopic observations was
therefore 
promptly started at ESO and CTIO. 
The complete data set will be presented elsewhere.

\section{Properties and evolution} \label{prop}
  
The SN was intrinsically faint compared to other SNe~II.  
If we assume a distance modulus of $\mu=30.64$ mag for NGC~1536
($H_0=75$
\kms\ Mpc$^{-1}$; \cite{tully}), and A$_B=0.00$ (RC3), a set of
closely 
spaced non-detections by Evans sets an upper limit to the SN
luminosity of 
\Mv(max) \gtsim $-14.65$.  This is about 2 magnitudes 
fainter than regular SNe~II at maximum (\cite{patII}).

In Fig.~\ref{lc_abs} we compare the absolute V light curve of \d\/
with 
those of other SNe~II: the double--peaked SN~1987A, the prototypical 
SNIIb 1993J and the linear SN~1979C. Even with the uncertainties on
the 
epoch of maximum and in the distance modulus, \d\/ was definitely
fainter than all of these objects. The decline rates in the V, R and
I
bands after J.D. 2450520 are close to the \co\/ decay rate.

The color of \d\/ at discovery was unusually red. 
No sign of interstellar absorption lines, which are expected in case
of 
strong extinction, is present in the spectra.  
\d\/ reddened rapidly over the first few weeks after discovery,
reaching 
a maximum \bv$=2.5$ in mid-February, and then turned gradually bluer.  
The analogy with the color curves of other SNe~II, which reach the
red-most 
point between 40 and 80 days after peak luminosity, indicates that
maximum 
light should have occurred around J.D. 2450430 ($\pm20$d).

A bolometric light curve is needed for a proper analysis of the
energetics.  
Unfortunately, most of our observations only cover the optical
domain.  
Only at one phase are nearly simultaneous optical (Jan.31) and IR
(Jan.27) 
spectra available.  The overall luminosity between 0.35 and 2.3 $\mu$
in 
the combined spectrum is log($L_{\rm bol}$) = 40.83 erg s$^{-1}$.  
At that epoch the color is still sufficiently blue that Schmidt's
(1998) 
empirical bolometric corrections to V can be used. One then obtains 
log($L_{\rm bol}$) = 40.96 erg s$^{-1}$, in good agreement with the 
spectrophotometry.  At later phases the color of
\d\/ was so red that Schmidt's correction cannot be applied. 
However, the available  spectra show that at all epochs most of the 
energy was radiated in the V and R bands and that the color evolution
is 
mostly due to variations in the B band, where the flux is much
smaller. 
Therefore, during the photospheric phase we assumed that the
bolometric 
light curve can be described reasonably well by applying a constant 
bolometric correction to the V magnitudes.  
The bolometric light curve thus derived is shown in the insert of
Fig.~\ref{lc_abs}.  The last point has been obtained adopting for
\d\/
the same $L_{\rm bol}/L_{\rm BVRI}$ ratio as for
SN~1987A in the nebular epoch (\cite{schm98}).

Although in SN~1987A the $\gamma$--rays from the \co\/ decay were not
completely deposited (\cite{pint}), the late time decline of SN~1987A
was close to the decay rate of \co. The mass of \ni\/ ejected by
SN~1987A was $\sim$ 0.075 \M\ (\cite{bouch91}, \cite{bouch91b},
\cite{bouch93}, \cite{danz}, \cite{nom94}). Assuming for \d\/ the
same $\gamma$--ray deposition as in SN~1987A, the ratio of the
luminosities on the tail of the light curves can be used to determine
the mass of \ni\/ synthesized in the explosion of \d. We then obtain
a ratio $M_{^{56}{\rm Ni}}$(87A)/$M_{^{56}{\rm Ni}}$(97D) $\sim 40$.
Taking into account the uncertainties in the determination of the
epoch of maximum and of the bolometric luminosity, the mass of \ni\/
ejected by \d\/ is very small, between 0.001 and 0.004 \M.

A comparison with SN~1987A (Fig.~\ref{97dvs87a}) shows general
similarities with respect to P-Cygni lines of \ion{H}{1},
\ion{Ca}{2}, \ion{Na}{1} and \ion{Ba}{2}, and more subtle
differences. These latter are emphasized by the strengths of the
\ion{Ba}{2} absorption lines and possibly those of other s--process
elements. SN~1987A showed strong lines of \ion{Ba}{2}, the analysis
of which (\cite{will87}, \cite{lucy88}, \cite{hoef88},
\cite{mazzchu}) suggested an enhanced abundance of barium. \d\/ shows
even stronger lines of \ion{Ba}{2}, with the identification of an
array of multiplets, but, as we shall discuss later, an overabundance
of barium is not required because a lower temperature prevails in the
envelope.


Strong lines of \ion{Sr}{2}, also an s--process element, were also
observed
in the IR spectrum taken on Jan.27 (\cite{cloc98}). In this spectrum
the 
\ion{He}{1} 10830\AA\/ line, which is the dominant feature in the IR
spectra 
of other SNe~II, is not clearly detected, although it might affect
the
blue wing of \ion{Sr}{2} 10914\AA.

Finally, one of the most striking peculiarities of \d\/ is the low
expansion velocity of the ejecta, approximately 1200 \kms, as
measured 
from the line absorption minima.  
Only the strong \ion{Ba}{2} lines show somewhat higher velocities, 
but it is well known that strong lines may form in outer parts of the 
envelope and that stratification effects may also be present.  
As we will discuss later (Sect. 3.2), the small velocity constrains
the 
epoch of the first spectrum to be $\sim 50$ days.

\section{Model Light Curve and Spectra} \label{mod_sect}

The light curve of SN~1997D is unique because the luminosity is low,
both at `maximum light' and on the radioactive tail.  These
properties, along with the narrowness of the lines and the redness of
the continuum, can be used to deduce the physical characteristics of
the progenitor and of the explosion.

The progenitor model used here is that of a 26 \M\ star with an 8
\M\/ He 
core at the beginning of collapse (\cite{nomha}) and an 18 \M\/
H-rich
envelope.    The stellar radius is
\r0\ = 300 \R, which is 3 - 5 times smaller than typical red
supergiants.  The explosion was simulated using a 1-D Lagrangian
hydrodynamical code (\cite{young}) with an energy of 
$E = 4 \times 10^{50}$ ergs and the ejection of 0.002 \M\/ of \ni.  
In order to eject a small amount of \ni, most of the material of the 
innermost region where Si is burned to Ni has to fall back on the 
degenerate neutron core where it is photodisintegrated before 
emitting $\gamma-$rays in forming the compact remnant. 
This is used to place the mass cut, resulting in a collapsing core 
of 1.8 \M, and ejecta mass of 24.2 \M. 

\subsection{Light Curves}

Our model light curve (Fig. \ref{lc_mod}) successfully reproduces the
main characteristics of the observations. 
The model parameters are constrained as follows.
The early light curve is powered by shock heating, i.e., the gradual
release of deposited shock energy.  
After the peak, the model light curve enters a slight plateau phase
lasting 
about 40 days (Fig.~\ref{lc_mod}); during this period, the
recombination 
wave moves inwards in mass through the H envelope down to the He
core.  
The peak luminosity ($M_V$ \gtsim --14.65) depends on \r0, $M$, and
$E$.  
In order to achieve a consistent model, the progenitor radius cannot
exceed 300 \R.  
For larger radii, the SN would have been too bright at shock
breakout.

After the plateau the light curve drops over 2 magnitudes to the
radioactive tail.  The luminosity and shape of the tail depend on the
\ni\/ mass and on $E/M$.  The \ni\/ mass determines the absolute
magnitude and $E/M$ determines the decline rate.  
The tail of SN~1997D follows the \co\/ decay rate as in SN~1987A.  
For the SN~1997D model, $E/M = 1.7 \times 10^{49}$ ergs \M$^{-1}$. 
This is $\sim$ 10 times smaller than in SN~1987A, and gives the low 
velocity and the complete trapping of the $\gamma$-rays. 
The low luminosity tail is reproduced with the low \ni\/ mass of
0.002 \M.

We have selected the current model through an elimination process
involving
7 models, which we will discuss in a subsequent paper.
For models with a small ejected mass (M $<$ 20 \M) the plateau is too
short, 
giving a time from explosion to discovery of $<$ 50 days,
inconsistent with 
the spectra. 
The tail of SN 1997D falls at the \co\/ decay rate, which suggests
two 
possibilities: 1) the ejecta is optically thick to the $\gamma$-rays
and the 
tail reflects the total \ni\/ mass ejected; or 2) the ejecta is
optically thin 
to the $\gamma$-rays and the tail is powered by annihilation of the
trapped 
positrons. Since the positrons carry only 3.7\% of the decay energy,
this 
latter case would imply a normal \ni\/ production. 
This can only be true if the opacity to $\gamma$-rays is less than
one,
i.e. if the total mass of the ejecta is
\begin{math}
M_{ej}  <  \frac{4  \pi  (v t)^2}{3  \kappa_{\gamma}}
\end{math}
. 
A simple estimate shows that for this to be the case the mass of the 
ejecta would have to be less than the mass of ejected \ni\/ alone. 
Using for the tail the values: $\gamma$-ray opacity, 
$\kappa_{\gamma} = 0.03 \ cm^2/g$, velocity, $v = 700$ km/s and 
time, t = 70 days, we get 
\begin{math}
M_{ej} < \frac{1}{80} $\M$
\end{math} 
. Thus the tail cannot be powered by positron annihilation and we
therefore 
explore models in which the ejecta is optically thick to
$\gamma$-rays.
Models with larger energies give higher material velocities and
broader 
spectral lines, which is again inconsistent with the observations.
For models with larger $E/M$ the light curve is too bright
and the plateau is too short.
Models with smaller progenitor radii have plateaus that are too
short. 
To attain the same final kinetic energy, the shock, in a more compact
star,
expands the ejecta, and thus cools it, much more rapidly than for a
star 
with a larger radius. The result is that the photosphere moves inward
faster, 
and the plateau is shorter.

\subsection{Spectra}

We used the explosion model described above to compute synthetic
spectra for
two epochs with a Monte Carlo spectrum synthesis code, which is based
on that
described in Mazzali \& Lucy (1993) but has been improved to include
photon
cascades (Mazzali \& Lucy 1998, in preparation).  

On Jan. 17 the red continuum indicates a photospheric temperature 
$T_{\rm eff} \sim 6400$ K.  The other input parameters are the
expansion 
velocity ($v_{\rm ph} = 970$ \kms ) and the epoch of explosion.  
Given the luminosity as deduced in Sect.~\ref{prop}, the required 
temperature can only be achieved with a rather large radius of about
6000 \R.  Since $R_{\rm ph} = v_{\rm ph} t$, this implies $t \sim 50$
days for the first spectrum. Therefore, we deduce that the peak was
missed by about 50 days, that a plateau followed, and that the SN was
first observed on the decline to the tail.

The model spectrum (Fig.~\ref{syn_sp1}) matches the observed one well
with the exception of some features in the red, where the model
continuum is also somewhat too high, and of the \Hb\ absorption. 
Most of the narrow absorption lines are well reproduced.  In
particular,
\ion{Ba}{2} 6497\AA\/ is stronger than \Ha. Since solar abundances
were
used, this is just a temperature effect, and no overabundance of Ba
is
required.  The temperature is low enough for the
\ion{Ba}{3}/\ion{Ba}{2} ratio to be small, which results in strong
\ion{Ba}{2} lines, while the same low temperature makes \Ha\/ weak. 
The narrow lines are clearly the result of a low explosion energy. 
The fact that the lines are so narrow allows a reasonably accurate 
determination of the mass above the photosphere. 
Values ranging between 10 and 20M$_{\odot}$ give acceptable spectra. 
The model adopted here has 14.5M$_{\odot}$ above 970 \kms. 
Models with a smaller mass give lines too shallow if $v_{\rm ph} =
970$ \kms.
 
Synthetic spectra show that the requirement 
that $t\sim 50$ days on 17 Jan. is a rather strong one.  
If the explosion took place closer to discovery, i.e. $t < 50$ days, 
the observed expansion velocity implies a small radius. 
Together with the derived luminosity, this results in a higher
$T_{\rm eff}$ than required.  
In this case, the \ion{Ba}{3}/\ion{Ba}{2} ratio increases and the 
\ion{Ba}{2} line intensities decrease.  
If on the other hand the explosion occurred earlier, so that $t > 50$ 
days, the radius is too large and the temperature becomes too low.  
When the temperature drops below 5000K, lines of \ion{Ti}{1} and
\ion{Fe}{1} absorb most of the radiation below 4500\AA\/ and the
synthetic spectrum changes dramatically.

We also computed a synthetic spectrum of our selected explosion
model evolved to match the 6 Feb. data. The fitting was satisfactory
using an epoch consistent with $t=50$ days on Jan.17.

\section{Discussion}

\d\/ is among the least luminous Type II supernovae known to date. 
The low luminosity persists also in the tail, where the decline rate
is 
close to that of a \co-powered tail.  \d\ is also characterized by a
low
temperature and a very low expansion velocity.  
These features are well reproduced by an explosion model with a
relatively 
small progenitor radius (300 \R), massive ejecta ($\sim$ 24 \M), low
explosion energy ($4 \times 10^{50}$ ergs), and small ejected 
\ni\/ mass ($\sim$ 0.002 \M).

A similarly small amount of \ni\ has recently been proposed
for another SN~II, SN~1994W (\cite{soll}), but this
object was rather luminous and its tail declined faster than the \co\
slope.

The neutrino-heating mechanism of massive star explosions is not well
understood, so the explosion energy and the amount of \ni\ ejected as
a function of progenitor mass are difficult to predict from the
hydrodynamical models.  
According to pre-supernova model calculations, stars more massive
than 
$\sim$ 25 \M\ form an Fe core in a greater gravitational potential
because 
of the significantly smaller C/O ratio and the consequently weak
carbon 
shell burning (\cite{nom93}; \cite{woowae}).  
Therefore, if the efficiency of neutrino heating does not change with 
exploding mass, more massive stars tend to produce lower explosion
energies, 
and they suffer from more fall-back of material onto the collapsed
remnant. 
The mass cut is then placed further out.  
However, the progenitor must be less massive than the Wolf-Rayet 
progenitors (i.e., \ltsim 30 - 40 \M) because of the presence of an 
envelope with a mass of at least 15 \M\ above the photosphere at day
50.

Stars in the range $\sim$ 10 - 25 \M\ are likely to explode with
typical explosion energies of $\sim 1 \times 10^{51}$ ergs, and to
produce 0.07 - 0.15 \M\ of \ni, as indicated by the brightness and
light curve shapes of SN 1993J, SN 1987A and of Type Ibc supernovae
(\cite{nom95}; \cite{young}).
Stars of 8 - 10 \M\/ are predicted to produce little \ni\ and other 
heavy elements (\cite{nom84}; \cite{maywil}).  
However, these stars are at the top of the AGB in the pre-supernova
stages 
(\cite{hashi}), which is inconsistent with the small radius of \d. 
Also, synthetic spectra obtained from the low mass models have
shallow 
absorption lines compared to the observations.

In conclusion, the progenitor of \d\ was probably as massive
as 25 - 40 \M.  Our 26 \M\ model is in this mass range, thus
being consistent with the low $E/M$ and the small \ni\ mass.  
The presupernova radius depends on many parameters, and the reason
for 
its being rather small in \d\/ is an open question.  Spectral models 
appear to rule out the case of low metallicity and He enhancement. 
A possible scenario involves a close binary system in which the
companion 
star spiraled in to make the envelope more massive (\cite{pod}).  

The ejection of a small mass of \ni\ provides a constraint on the
mass
of the collapsed remnant, $M_{\rm rem}$ (\cite{nom93}; \cite{thiel}).  
In our 26 \M\ model, is $M_{\rm rem} = 1.8$ \M. 
If $M_{\rm rem}$ were assumed to be smaller, more \ni\ would be
ejected:  
e.g. if $M_{\rm rem} = 1.7$ \M, then M(\ni) = 0.1 \M.
For more massive progenitor models, $M_{\rm rem}$ would be larger
than 1.8 \M. 
If the equation of state of nuclear matter is relatively soft, the
maximum
(gravitational) mass of a cold neutron star is $\sim$ 1.5 \M\
(\cite{bb94}).  
If this is the case, the remnant of \d\ must be a small-mass black
hole, 
and the low luminosity tail of \d\ might be a signature of the
formation 
of such an object.

\bigskip
We would like to thank Piero Rosati for observing \d\/ and Rev.
Robert
Evans for providing his pre-discovery limits.  P.A. Mazzali
acknowledges receipt of a Foreign Research Fellowship at N.A.O., and
is grateful to T. Kajino and the Dept. of Astronomy at the University
of Tokyo for the hospitality.  A. Piemonte aknowledges financial
support from ESO during his stay in Chile.  This work has been
supported in part by the Grant-in-Aid for Scientific Research
(05242102, 06233101) and COE research (07CE2002) of the Ministry of
Education, Science, and Culture in Japan, and by the National Science
Foundation in US under Grant No. PHY94-07194.  Part of the
observations have been obtained at the ESO 1.5m telescope operated
under the agreement between ESO and Observatorio Nacional -- Brasil.

\newpage

 \figcaption[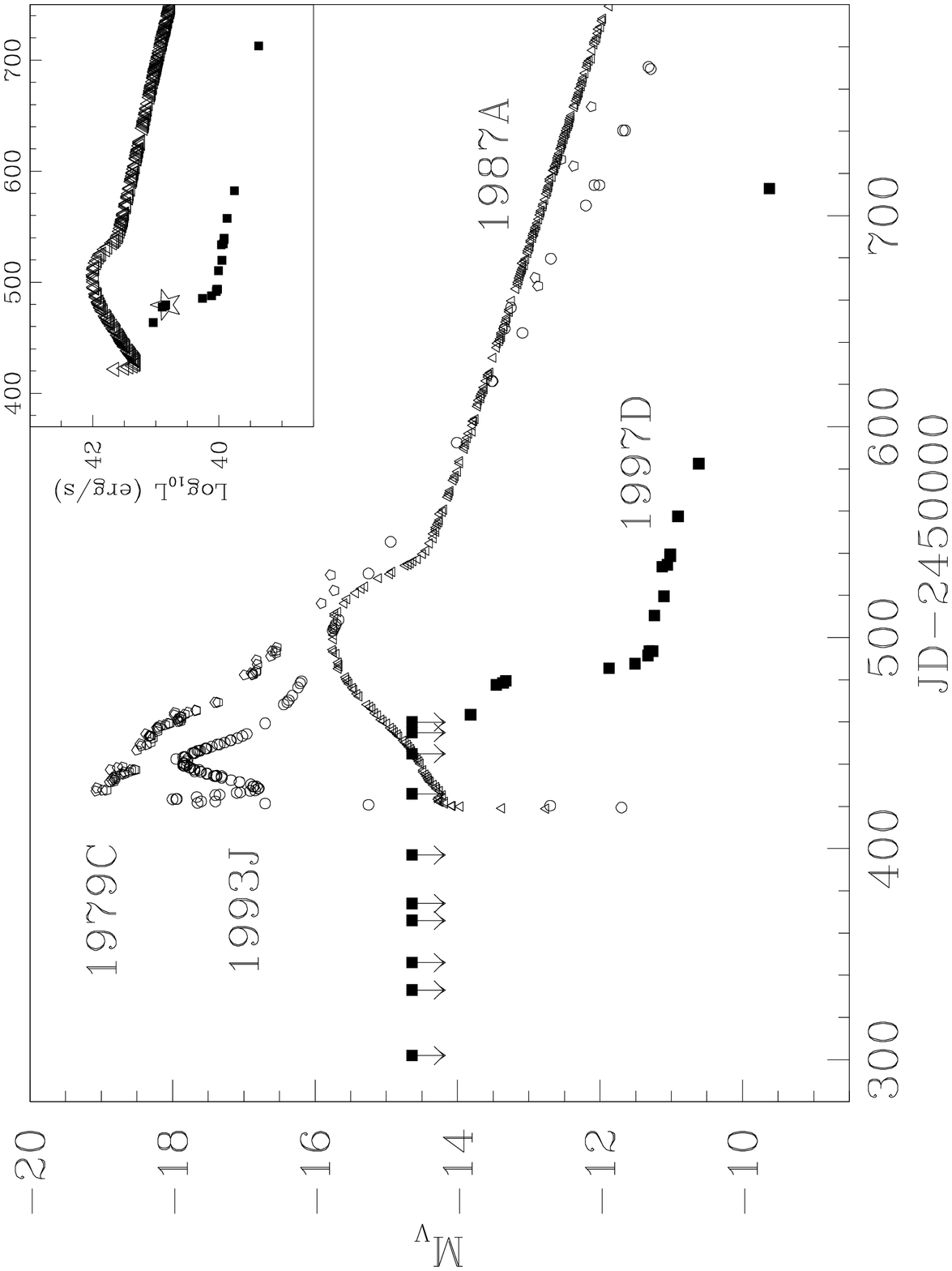]{Comparison of the absolute V 
light curve of SN~1997D with those of SNe 1987A, 1993J and 1979C.  
The epochs of maxima have been corrected to J.D. 24500420.  
Only galactic absorption has been considered. 
The upper limits from the negative observations of Rev. Evans
(private 
communication) have been also plotted.  
The insert shows the bolometric light curves of \d\/ and SN~1987A 
(Catchpole et al. 1987, 1988). 
The star marks the optical+IR spectrophotometric observation of
Jan.30.  
\label{lc_abs}}

 \figcaption[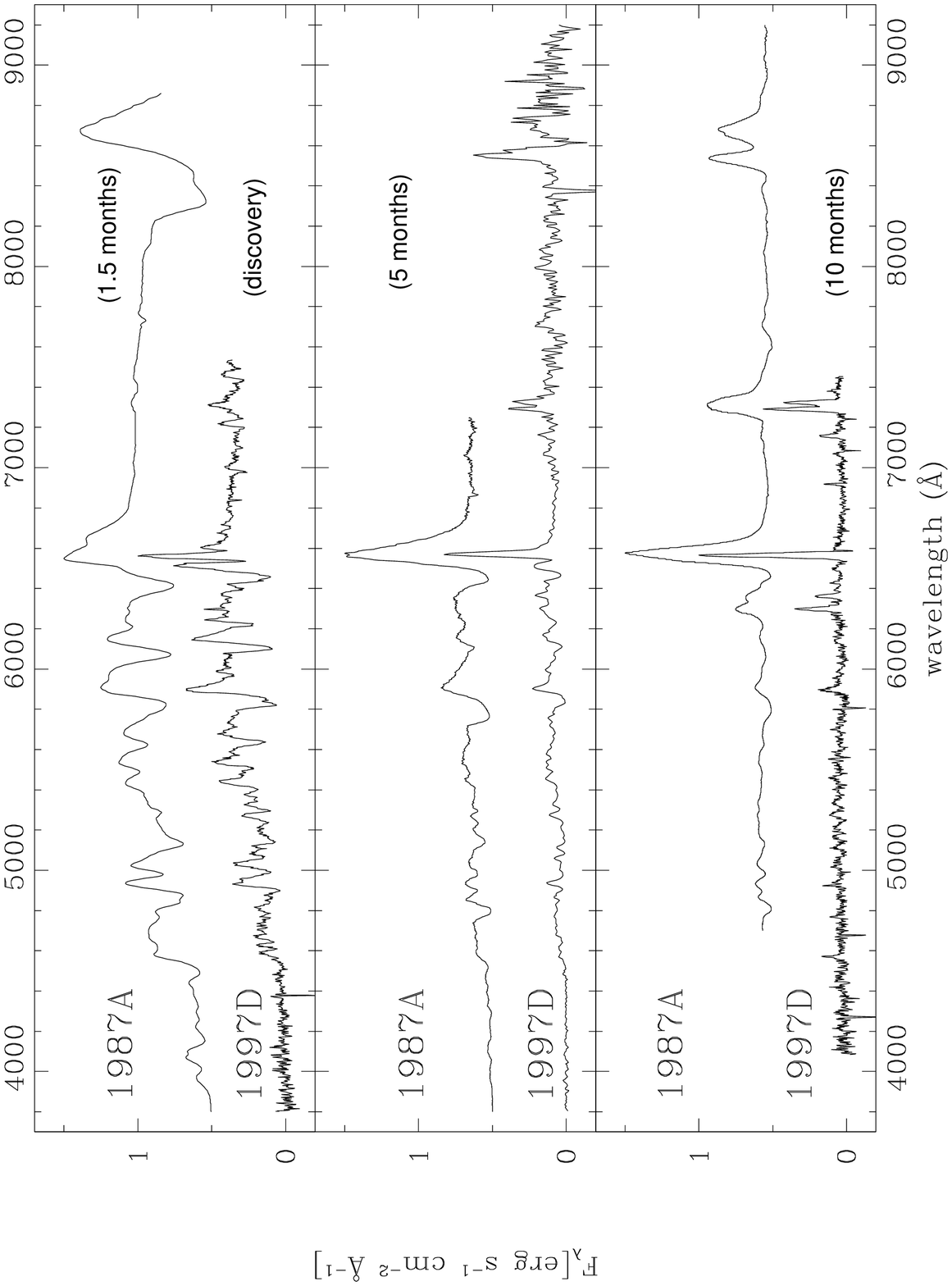]{Comparison of the spectra of
 SN~1997D and SN~1987A.  
 Top: the spectrum of SN~1997D at discovery is compared with that of 
 1987A 1.5 months past explosion.
 Middle: spectra taken 100 days later. The spectrum of SN~1997D is
 the
 result of the merging of two spectra with different resolutions and
 signal--to--noise ratios. 
 Bottom: spectra taken 10 months past maximum.  
 All spectra have been normalized to the peak of the \Ha\ emission
 and 
 corrected for redshift.
\label{97dvs87a}}

\figcaption[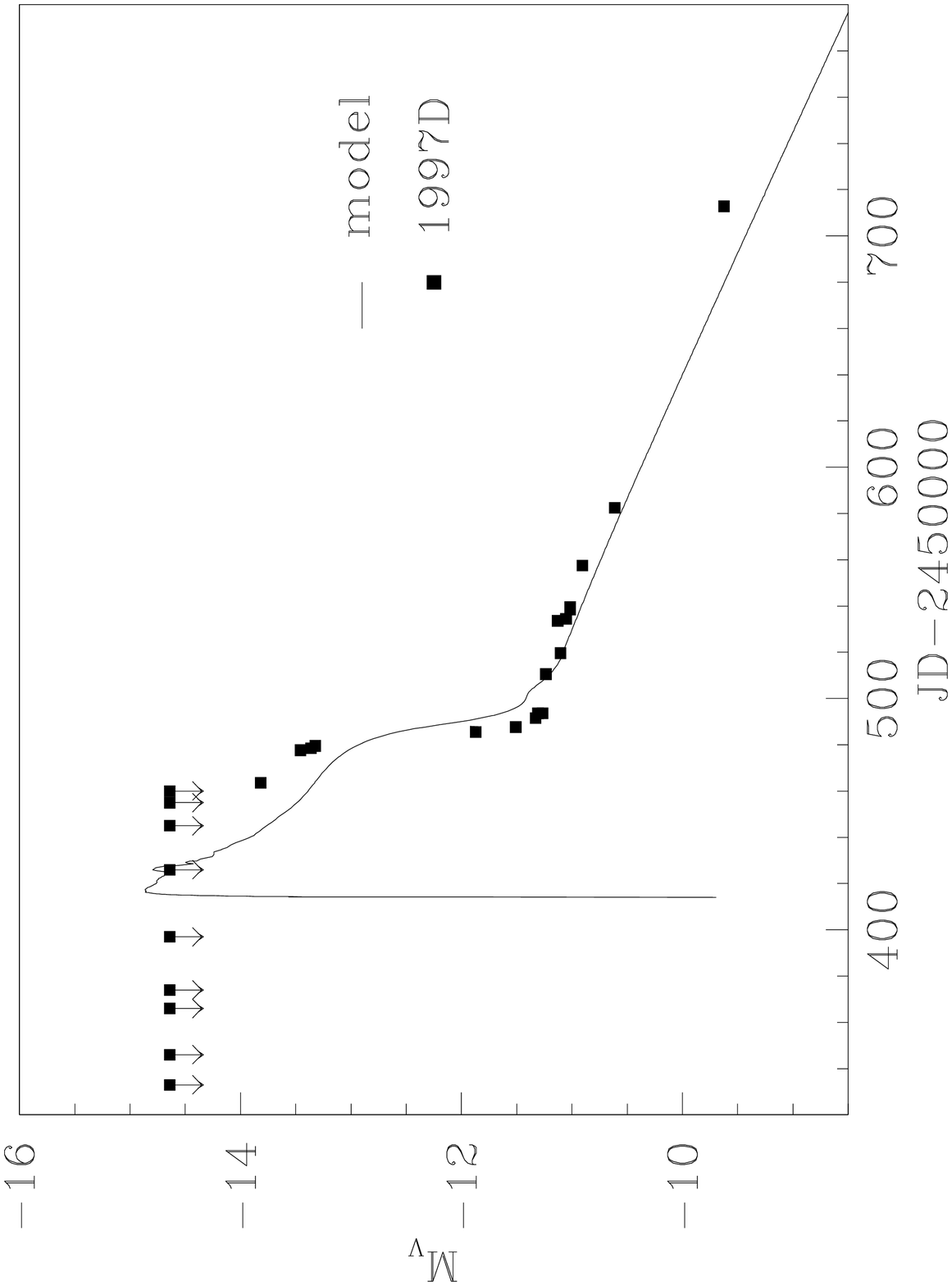]{The observed absolute V photometry of
\d\/ 
 compared to the model light curve (mass $=24$ \M; radius $=300$\R;
 mass of \ni$=0.002$ \M; energy$=4 \times 10^{50}$ ergs). 
\label{lc_mod}}

\figcaption[./fig/sp_mod2_97d.ps]{The observed spectrum of \d\/ on
Jan.17 
 (continuous line) and the synthetic spectrum computed from the
 explosion 
 model (dashed line). 
 The parameters of the model are: $v_{\rm ph}=970$ \kms;
 radius$=6000$ \R;
 $T_{\rm eff}= 6400$ K; mass above the photosphere $M=14.5$ \M;
 epoch=50 days; solar abundances. 
\label{syn_sp1}}

\end{document}